\documentclass{astlb} 
\usepackage[english]{babel}
\usepackage[utf8]{inputenc} 
\usepackage{amsmath}
\usepackage{amssymb}                                          
\usepackage{graphicx}

\usepackage{mathtext}
\usepackage[hyperfootnotes=false]{hyperref}
\usepackage{color}
\usepackage{epsf}
\usepackage{amsfonts}
\usepackage{gensymb}
\usepackage{float}
\usepackage{ulem}
\usepackage{diagbox}
\setcitestyle{numbers}

\def\tightleading{1.1}
\def\doubleleading{1.6}
\def\baselinestretch{\doubleleading}

\let\tightenlines=\tighten

\tightenlines

\def\apj{ApJ}
\def\apjl{ApJL}
\def\apjs{ApJ Suppl. Ser.}

\def\aap{A\&A}

\def\araa{ARAA}
\def\mnras{MNRAS}

\def\nat{Nature}
\def\pasj{PASJ}
\def\pasa{PASA}

\begin{document}
%\baselineskip 21pt

%\journalinfo{2017}{0}{0}{1}[0]
%\UDK{524.77}

%\title{Can observations of 511 keV line from nearby galaxies shed light on the AGN jet composition?}
\title{Can observations of 511 keV line from the M31 galaxy shed light on the AGN jet composition?}
\author{B.A. Nizamov$^{1}$\email{nizamov@physics.msu.ru}, M.S. Pshirkov$^{1,2}$,   \\
$^1$ Sternberg Astronomical Institute, Lomonosov Moscow State University, Universitetsky pr., 13, Moscow, 119234, Russia\\
%$^2$ Institute for Nuclear Research of the Russian Academy of Sciences (INR RAS), 
%117312, Москва, проспект 60-летия Октября, 7а\\
$^2$ Lebedev Physical Institute, Pushchino Radio Astronomy Observatory
}

\submitted{00.00.2018}

%\maketitle
\pagebreak
\begin{abstract}

%\section*{Abstract}
Positron annihilation line at 511~keV is a known component of the gamma-ray diffuse emission. It is believed to be produced in the Galaxy, but there could be possible extragalactic contribution as well. E.g., positrons can be produced in jets of active galactic nuclei (AGN) and after that accumulate and gradually annihilate in hot gaseous halos around galaxies. In this work we test this hypothesis in application to an individual object -- the Andromeda galaxy (M31) which is close and has a supermassive black hole in its center, which powered an AGN before. We compute the growth history of the supermassive black hole in M31, relate it to the evolution of jet luminosity and estimate the positron content in its halo. We calculate the 511~keV photon flux due to positron annihilation which should be observed at Earth and find the value of around $10^{-4}$ photon cm$^{-2}$s$^{-1}$. It is very close to the observational limits ($<10^{-4}$photon cm$^{-2}$s$^{-1}$) set by the INTEGRAL/SPI in the assumption of the point source, so further observations would be able to constrain leptonic models of the jets and propagation of cosmic rays in the circumgalactic medium of large spiral galaxies.

\englishkeywords{astroparticle physics -- galaxies: active -- galaxies: jets -- gamma-rays: galaxies}
\end{abstract}
%\pagebreak

%\pagebreak 

%%%%%%%%%%%%%%%%%  INTRO %%%%%%%%%%%%%%%%%%%%%%%%%
\section{Introduction}
Super-massive black holes (SMBHs) in active galactic nuclei (AGNs) could launch powerful relativistic jets during accretion phases. These jets carry energy in particles and fields  from the central engine to the surrounding medium. The exact particle composition of jets is not perfectly known: it could consist either of  ions (mainly protons) and electrons, where number of protons and electrons are close, or, alternatively, jet could be pair-dominated, i.e. with a large fraction of $e^+e^-$ pairs and $n_{e^+}\approx n_{e^-} \gg n_{p} $, where $n_{e^+}, n_{e^-},n_{p}$ are number densities of positrons, electrons and protons respectively.

Relativistic leptons -- both electrons and positrons--actively participate in radiative processes in the jet via  synchrotron  and inverse Compton emission in magnetic fields and background photon fields correspondingly, thus allowing us to observe the jet in different frequency ranges from radio to gamma. Eventually, these leptons make their way into surrounding medium -- interstellar (ISM) and, later, circumgalactic (CGM).

If the jets are pair-dominated then the AGNs could possibly be one of the main sources of positrons in the Universe. On much smaller scales it could also be the case in the Galaxy, where jets of microquasars, powered by accretion on stellar mass BHs could produce significant or even the major fraction of the galactic positrons \cite{Heinz2002,Siegert2016}.

The pair content has been studied in literature via modelling of
the radiative properties of jets. Zdziarski et al. \cite{Zdziraski2022MAXI}, \cite{Zdziarski2022Cyg}, \cite{Zdziarski2022galaxy} used synchrotron-self Compton spectral fits to deduce the electron flow in the jet and hard
X-ray data to find the pair production rate in its base, and the two quantities appeared to correspond to each other. In the works \cite{Sikora2016} and \cite{Pjanka2017} the pair content was estimated from the overall energetics of jets. A more direct observational test was suggested by Ghisellini \cite{Ghisellini2012} who proposed that pair-abundant jet bases should have enhanced brightness at $\sim 1$~MeV even for misaligned sources. An interesting scenario is discussed in \cite{Vecchio2013}: if positrons are produced in powerful extragalactic sources such as AGNs and escape to intergalactic medium, they can survive there practically infinitely until they accrete on to galaxies, e.g. Milky Way. The authors do not estimate the production rate of positrons, but they find that a positron to electron density ratio in the Universe of $10^{-5}$ is sufficient to explain the 511~keV diffuse emission in the Galaxy. In the present work, we propose quite a straightforward test of pair production via estimation of the 511~keV flux. 
More concretely, we study the ultimate fate  of the positrons after they leave the jet. Observations show that the Milky Way-size galaxies, i.e. $M_*\sim10^{11}~M_{\odot}$ are surrounded by vast tenuous halos where density gradually decreases to $\sim10^{-4}~\mathrm{cm}^{-3}$ at galactocentric radii $r\sim(50-70)$~kpc \cite{Tumlinson2017}. This CGM has a very complex multiphase structure, where denser and colder regions with $T\sim10^{5}$~K are surrounded by hotter and more diluted plasma with temperature closer to virial $T\sim10^{6}$~K.
We assume the following scenario: positrons brake and are trapped in the extended CGM of the host galaxy.  Time scales of braking and thermalization with the plasma of the CGM depends on the density, after that annihilation begins to operate effectively.\footnote{For the realistic properties of leptons and halo the energy losses before thermalization due to in-flight direct annihilation amount only to several \%  of the total energy. E.g., in \cite{Beacom2006} it is shown that, in the Galaxy, positrons injected with the Lorentz factors 20, 6, 2 lose, respectively, 11, 5.5, 1.4\% of their energy before they thermalize. In our work, we will assume $\Gamma \leq 20$ (see Section~\ref{sec:discussion}), therefore we will neglect the energy lost via in-flight annihilation.}
The annihilation could proceed either through direct in-flight channel, $e^{+}+e^{-} \longrightarrow 2\gamma, ~E_{\gamma}=511$~keV or through the bound state, so-called positronium (Ps). Ps could form in two states, depending on mutual orientation of $e^{+}$ and $e^{-}$ spins: the singlet one, para-positronium (p-Ps) in $\sim10^{-10}$~s decays in 2 photons with $E_{\gamma}=511$~keV; the triplet, ortho-positronum (o-Ps) lives longer, $\sim10^{-7}$~s and decays into three gamma photons, forming a continuum. Total branching ratio of  p-Ps and o-Ps formation is 1:3.
Annihilation time scales depend on density and temperature of halo gas and the initial positron energy and appear to be extremely long ($>t_\mathrm{br}$) as we show in Sec.~\ref{sec:braking}. Still if the positrons are effectively retained in the halo (see, e.g., \cite{Crocker2011, Feldmann2013, Lacki2015, Hopkins2020} for case of CR protons) for Gyrs, we could expect emergence of 511 keV annihilation line as a smoking gun for pair-dominated jets. An analogous idea was previously put forward in \cite{Furlanetto2002}, but in that work the authors assumed that the positrons are ejected to the intracluster medium. Its high temperature (above $10^7$~K) suppresses positronium formation so that effectively all annihilations result in 511~keV line emission.
In this paper we focus on the constraints that could be obtained from the observations of individual galaxies in the local Universe. Constraints from the cosmological signal, produced by the possible background from red-shifted annihilation lines from the halos of all galaxies would be studied elsewhere.
Sgr~A$^*$ as a source of positrons was considered in \cite{Totani2006, Jean2017}. The difference is that these papers were studying present time  ($\sim$~Myr) positron production and annihilation in the galactic ISM, while  we are dealing with much longer time scales ($\sim$~Gyr) and annihilation in the much more extended halo.

%%%%%%%%%%%%%%%%  MAIN PART  %%%%%%%%%%%%%%%%%%%%%%%%%%%
\section{511~keV flux estimation}
The radiation at 511~keV is produced by the annihilation of electron-positron pairs. In our approach,
there are several assumptions which provide physical ground for the whole calculation. We first outline them
briefly and then discuss in more detail in the following subsections.

1) Growth history of the SMBH can be derived from the luminosity function of AGNs at various redshifts via the continuity equation, if some relation between the AGN luminosity, accretion rate and accretion efficiency is assumed (Section~\ref{sec:growth}).
Due to this relation, we obtain not only the SMBH growth history, but also the AGN luminosity history.

1) Positrons are born in the jet of AGN. We assume that $n_\mathrm{pair}\sim 15$ positrons are produced per proton (Section~\ref{sec:production}).

2) We suppose that the host galaxy is surrounded by the halo with the density $n \sim 10^{-4}$~cm$^{-3}$. For the temperature we consider two bracketing cases, i.e. $T=10^{5}$~K and $T=10^{6}$~K. We expect that results for real multiphase system would lie in between them. In this halo, positrons brake due to Coulomb collisions and thermalize. Subsequently, they annihilate
with the medium electrons directly or via positronium formation. We assume that the positrons are retained in the CGM for cosmological times.

The final result depends on several parameters, such as the Eddington ratio, accretion efficiency, bulk Lorentz factor of
the jet, the halo density. We choose values which are preferred observationally or have some theoretical basis in the literature.

%%%%%%%%%%%%%%%%  SMBH GROWTH  %%%%%%%%%%%%%%%%%%%%%%%%
\subsection{SMBH growth and AGN luminosity} \label{sec:growth}
Since the famous work of Soltan \cite{Soltan1982}, there has been many attempts to relate the growth history of SMBHs with the evolution of the luminosity function of AGNs hosted by them. In our work, we follow Marconi et al. \cite{Marconi2004} (hereafter M04). For the paper to be more self-contained, we briefly repeat the ideas of this paper relevant to us.

If $N(M, t)$ is the comoving number density of SMBHs with the mass $M$ at the cosmic time $t$ and $\langle\Dot{M}\rangle$ is the average accretion rate of a SMBH with the mass $M$, then the continuity equation holds:
\begin{equation}
    \frac{\partial N(M, t)}{\partial t} + \frac{\partial}{\partial M} \left[N(M, t) \langle\Dot{M}(M, t)\rangle \right] = 0. \label{eq:continuity}
\end{equation}
The AGN luminosity function can be related to the mass function as follows:
\begin{equation}
    \phi(L, t)d \log L = \delta(M, t) N(M, t)dM \label{eq:phi}
\end{equation}
where $\delta(M, t)$ is the fraction of SMBHs active at time $t$.

\begin{figure}
    \centering
    \includegraphics[scale=0.8]{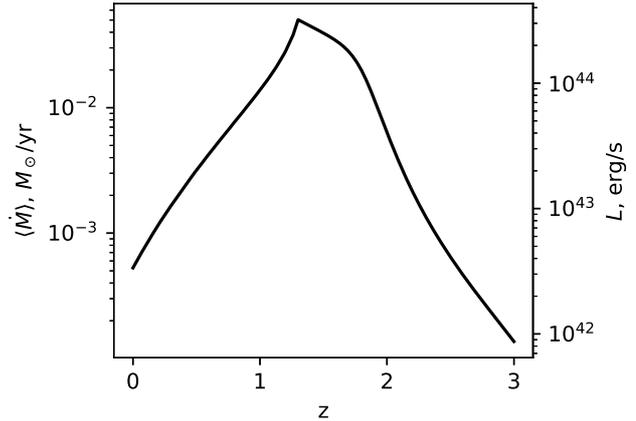}
    \caption{Average growth rate of an SMBH with the initial mass $1.56 \times 10^4 M_\odot$. On the right axis is shown the corresponding average AGN bolometric luminosity.}
    \label{fig:mdot}
\end{figure}

The accretion efficiency $\varepsilon$ is defined as the fraction of the infalling matter rest energy which is converted into radiation, and the Eddington ratio $\lambda$ is the ratio between AGN luminosity and Eddington luminosity. With these definitions, the luminosity, mass and accretion rate are related as follows:
\begin{equation}
    L = \lambda\frac{Mc^2}{t_\mathrm{Edd}} = \varepsilon \Dot{M}_\mathrm{acc} c^2 \label{eq:L}
\end{equation}
where $t_\mathrm{Edd}$ is the Eddington time and $\Dot{M}_\mathrm{acc}$ is the accretion rate. The BH growth rate then
equals $\Dot{M} = (1 - \varepsilon)\Dot{M}_\mathrm{acc}$. The quantity $\langle\Dot{M}\rangle$ is the growth rate averaged over the whole population of BHs of mass $M$, in other words, $\langle\Dot{M}\rangle = \delta(M, t)\Dot{M}$. From Eqs.~\ref{eq:phi}-\ref{eq:L} follows
\begin{equation}
    N(M, t) \langle\Dot{M}(M, t)\rangle = \frac{1-\varepsilon}{\varepsilon c^2 \ln{10}} \phi(L, t)_{L=\lambda Mc^2/t_\mathrm{Edd}} \frac{\mathrm{d}L}{\mathrm{d}M}.
\end{equation}
This can be substituted to Eq.~\ref{eq:continuity} and, with $\varepsilon$ and $\lambda$ constant, we obtain
\begin{equation}
    \frac{\partial N(M, t)}{\partial t} = -\frac{(1-\varepsilon)\lambda^2 c^2}{\varepsilon t_\mathrm{Edd}^2 \ln{10}} \left[\frac{\partial \phi(L, t)}{\partial L}\right]_{L=\lambda Mc^2/t_\mathrm{Edd}}. \label{eq:N}
\end{equation}
For an initial condition, M04 assumed that at the starting redshift $z_s=3$ all the SMBHs were active, i.e. $\delta(M, t(z_s)) = 1$ which implies
\begin{equation}
    MN(M, t_s) = [\phi(L, t_s)]_{L=\lambda Mc^2/t_\mathrm{Edd}}. \label{eq:initial}
\end{equation}
Equations~\ref{eq:N}, \ref{eq:initial} can be solved once we have data on the AGN luminosity function from sufficiently large
$z$ up to the present. Like M04, we used a representation of $\phi(L, t)$ by Ueda et al. \cite{Ueda2003}.
%https://latex.org/forum/viewtopic.php?t=14231
From Eq.~\ref{eq:N} we get an expression for the average growth rate:
\begin{equation}
    \langle\Dot{M}(M, t)\rangle = \frac{1}{t_\mathrm{Edd} \ln{10}} \frac{(1-\varepsilon)\lambda}{\varepsilon N(M, t)} \phi(L, t)_{L=\lambda Mc^2/t_\mathrm{Edd}}.
\end{equation}
In Fig.~\ref{fig:mdot} we show how $\langle\Dot{M}(M, t)\rangle$ evolves. On the right axis of this plot we also show
the average bolometric luminosity which is obtained from Eq.~\ref{eq:L} if we replace $\Dot{M}(M, t)$
with $\langle\Dot{M}(M, t)\rangle$.

We also need to select  candidates that could provide the strongest  constraints. It is obvious that the stronger signal would be expected from the nearest galaxies with more massive black holes. There are three possible candidates -- the Milky Way ($M_\mathrm{SMBH}=4\times10^{6}~M_{\odot}, d\sim 50~\mathrm{kpc}$), where we use the characteristic halo radius as a source distance \cite{Ghez2008}, the M31 galaxy ($M_\mathrm{SMBH}\sim(1-2)\times10^{8}~M_{\odot},~ d\sim 750~\mathrm{kpc}$,\cite{Bender2005}), the Cen~A galaxy ($M_\mathrm{SMBH}=5.5\times10^{7}~M_{\odot},~d\sim 3.5~\mathrm{Mpc}$, \cite{Neumayer2010}). M31 seems to be the perfect candidate -- the expected total flux is comparable to one from the Milky Way. However, the signal from the halo of our own galaxy is almost uniform, while from the M31 galaxy it would be much more localized with the angular size around $10^{\circ}$.

\begin{figure}
    \centering
    \includegraphics[scale=0.8]{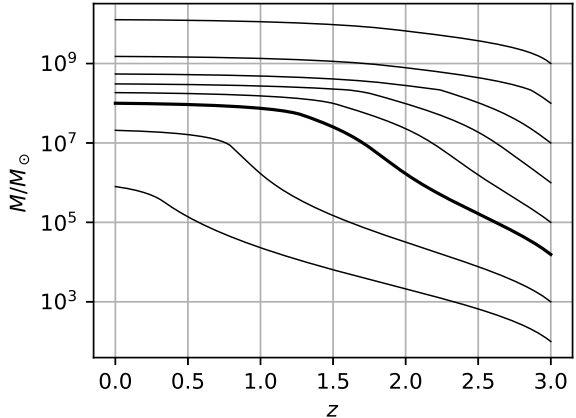}
    \caption{Growth history for SMBHs of various initial masses calculated by method of \cite{Marconi2004}. The track
    used in the calculations is shown in bold. }
    \label{fig:tracks}
\end{figure}

Now we can choose a starting mass of a BH at $z_s = 3$ and integrate the equation $dM = \Dot{M}(M, t) dt$ to obtain the
BH mass at $z=0$. In particular, we can find by trial such a starting mass that finally gets to the present mass of the
SMBH in M31. Different estimations of this mass fall into the range $5\times 10^7 - 1.4\times 10^8 M_{\odot}$ \cite{Menezes2013}, we adopt the value of $10^8 M_{\odot}$. We checked ourselves by calculating a number of tracks for different starting BH
masses and obtained a plot similar to Fig.~8 of M04, it is shown in Fig~\ref{fig:tracks}. The track shown in bold starts
at the mass of $1.56 \times 10^4 M_\odot$ and ends with $10^8 M_\odot$,  we use it in our calculations as a proxy of M31  SMBH growth history.

Like M04, we adopt the value of the accretion efficiency $\varepsilon=0.1$ and the Eddington ratio $\lambda=1$. Furthermore, we assume that the
jet kinetic power is tightly connected to the accretion rate \cite{Ghisellini2014}:
\begin{equation}
    P_\mathrm{j} = \eta \Dot{M}_\mathrm{acc} c^2,~\eta\sim 1 \label{eq:pjet}
\end{equation}
The kinetic power is close to the accretion power, as also shown in \cite{Ghisellini2014}. Note that the estimate that the factor $\eta$ is close to unity was obtained in \cite{Ghisellini2014} in assumption that there was one proton per electron in the jet. Although the authors argue for the presence of protons in the jet, the number of protons per electron is not a parameter of the model used, as stated in \cite{Ghisellini2009}. If the actual proton load of the jet is smaller,  i.e. there are more than one electron per on proton,  then the total jet power is also smaller (see discussion before Eq.~\ref{eq:pjet2}).

%%%%%%%%%%%%%%%%%  POSITRON PRODUCTION  %%%%%%%%%%%%%%%%%%%%%%
\subsection{Positron production} \label{sec:production}
The composition of AGN jets is currently a matter of investigation. In \cite{Ghisellini2012}  it is argued that pairs can be created
in the inner part of the jet due to photon-photon collisions if the luminosity is sufficient, i.e. above $10^{44}$~erg/s
at 1~MeV. The discrepancy between jet powers calculated from blazar spectral fits and
from radio lobe calorimetry was discussed  by Sikora in  \cite{Sikora2016}. He investigated non-zero pair content as a cause for the power underestimation by spectral
fits and came to conclusion that presence of about  15 pairs per proton can reconcile estimations by the two methods. Similar
question was studied in \cite{Pjanka2017}. The authors compared jet powers estimated by spectral fitting,
radio-core shift, radio lobes and a phenomenological estimate based on gamma-ray luminosity. They found that on average
spectral fitting and core shift method give powers which are ten times larger than that from radio lobes method. Like in \cite{Sikora2016}, they suggest the presence of $\sim 15$ pairs per proton as a possible explanation
which would reduce the power estimated from spectral fitting.

In a recent series of papers, Zdziarski et al. investigated the composition of jets in two black hole X-ray binaries
MAXI~J1820+070 \cite{Zdziraski2022MAXI} and Cyg~X-1 \cite{Zdziarski2022Cyg} and in a radio galaxy 3C~120
\cite{Zdziarski2022galaxy}. In particular, they estimated the pair production rate at the jet base of 3C~120 and it
appeared to correspond quite well to the flux of synchrotron emitting electrons downstream in the jet. They also found
that the kinetic power in ions greatly exceeds the maximum possible jet power if ions are equally abundant as
electrons. Interestingly, they arrive at similar conclusions in the case of two X-ray binaries.

One can notice that in the works dealing with jets in AGN, the estimates of $n_\mathrm{pair}$ is approximately 10--20.
We can adopt this value in our calculations, but we show later that the exact value does not affect our conclusions.
We only suppose that positrons are produced in a certain amount compatible with observations.

The total jet power is given by Eq.~\ref{eq:pjet}. As we already mentioned, this estimation is obtained in \cite{Ghisellini2014} in the assumption that there is one proton per electron in the jet. They also found that the total power of the jet appears to be dominated by protons: their power is, on average, more than an order of magnitude larger than the  radiation power -- the second-largest component. The total number of leptons is constrained by the observed amount of radiation. The number of protons could be obtained from that, given that we know the ratio of the number of protons to the number of leptons. Therefore, if we allow for positrons in the jet, the number of protons is reduced accordingly. Since the jet power is dominated by protons, it will diminish proportionally. In particular, if there are $n_\mathrm{pair}$ positrons per proton then the jet power is diminished by a factor $2n_\mathrm{pair}$:
\begin{equation}
    P_\mathrm{j} = \eta \Dot{M}_\mathrm{acc} c^2 / 2n_\mathrm{pair}. \label{eq:pjet2}
\end{equation}
As this power is supplied by protons, we can relate the jet power to the rate of the proton flux $\Dot{N}_\mathrm{p}$ (i.e. the number of protons traversing the jet cross-section per unit time) via the jet bulk Lorentz factor $\Gamma$ for which we choose the benchmark value of 10:
\begin{equation}
    P_\mathrm{j} = \Dot{N}_\mathrm{p} \Gamma m_\mathrm{p} c^2. \label{eq:pjet3}
\end{equation}
The positron production rate equals $\Dot{N}_+ = n_\mathrm{pair} \Dot{N}_\mathrm{p}$. Combining this with
Eqs.~(\ref{eq:pjet2}-\ref{eq:pjet3}) we obtain
\begin{equation}
    \Dot{N}_+ = \frac{\eta \Dot{M}_\mathrm{acc}}{2 \Gamma m_\mathrm{p}}, \label{eq:ndot}
\end{equation}
for further numerical estimates we will assume benchmark value $\eta=1$.
One can see that $n_\mathrm{pair}$ is cancelled out. This is natural in our setup: with larger $n_\mathrm{pair}$ more pairs
are produced, but the jet kinetic power estimated from the electron content decreases, and both dependencies are linear.
We emphasize that the above formulae are valid if the jet power is dominated by protons. This may be not the case when $n_\mathrm{pair} \gtrsim 10$. However, such a scenario is rather questionable because, as stated in \cite{Ghisellini2014}, it would mean the jet power is less than the radiation power and the jet would stop.
%%%%%%%%%%%%%%%%%%  BRAKING & ANNIHILATION  %%%%%%%%%%%%%%%%%%%%%%
\subsection{Positron braking and annihilation}\label{sec:braking}
Positrons produced in the jet are (at least) moderately relativistic. To annihilate efficiently, they should thermalizae
in the ambient medium. The medium is the halo gas with temperature of order $10^5 - 10^6$~K and density
$10^{-4}-10^{-3}$~cm$^{-3}$. This gas is ionized, therefore positrons brake mostly due to Coulomb collisions with ions.
For the energy loss rate we take Eq.~(14) from \cite{Prantzos2011} and estimate the brake time as the initial positron
kinetic energy divided by the loss rate:
\begin{equation}
    t_\mathrm{br} = (\Gamma-1) mc^2 \left\{ 7.7 \times 10^{-9}\frac{n}{\beta}\left[\ln\left(\frac{\Gamma}{n}\right) + 73.6 \right]\right\}^{-1}
\end{equation}
where the expression in braces is in eV/s, $n$ is the halo density in cm$^{-3}$, $\beta$ is the positron velocity in units of $c$.
Positrons are not expected to be monoenergetic. However, if their spectrum is not too hard, the overwhelming majority of the particles  is
contained in the low-energy part of the spectrum, therefore we assume that all the positrons initially have the same Lorentz
factor as the jet.
%For $\Gamma=10$ and $n=10^{-4}$~cm$^{-3}$ we get $t_\mathrm{br} = 2.2$~Gyr.
For the parameter range of interest, the above equation can be written as
\begin{equation}
    t_\mathrm{br} \approx 2.2 \times (\Gamma/10) (n/10^{-4}\text{cm}^{-3})^{-1}\quad \text{Gyr.} \
\end{equation}

Also the leptons could  brake down more efficiently, if they lose their energy trough adiabatic cooling. It could be the case if they stayed attached to the expanding galactic wind rather than simply diffuse away in the outer halo. In this case the characteristic time scale could be estimated as $t_{ad}\sim r_{halo}/v_{wind}$, and for benchmark values of $r_{halo}=50~$kpc, $v_{wind}=300 \mathrm{km/s}$ this time scale would be around 200 Myr.
Positrons can annihilate directly or after formation of positronium, the rate constants of the two processes being respectively $\langle \sigma_\mathrm{a} v \rangle$ and $\langle \sigma_\mathrm{r} v \rangle$ where subscripts stand for "annihilation" and "recombination". We calculate these rates according to the formulae from \cite{Gould1989}. For the halo density $n \sim 10^{-4}$~cm$^{-3}$ and temperature $T \sim 10^6$~K the time of annihilation appears to exceed the braking time. Again, for reasonable $T, n$ we can roughly estimate these times as
\begin{eqnarray}
t_\mathrm{a} \approx 20 \times (T/10^6 \text{K})^{0.5} (n/10^{-4}\text{cm}^{-3})^{-1}\quad \text{Gyr} \\
t_\mathrm{r} \approx 27 \times (T/10^6 \text{K})^{1.1} (n/10^{-4}\text{cm}^{-3})^{-1}\quad \text{Gyr}
\end{eqnarray}
 To obtain the total annihilation rate in the halo, we must calculate the positron content there which is determined by the balance between the positron production and annihilation. We neglect annihilation 'in-flight', i.e. before thermalization, therefore what we need is the number of \textit{thermalized} positrons. Their production rate at time $t$ equals $\Dot{N}_+$ from Eq.~\ref{eq:ndot} at time $t-t_\mathrm{br}$ because newly created positrons need $t_\mathrm{br}$ to brake. 
To sum up, we solve the following equation
\begin{equation}
    \frac{d N_+(t)}{dt}= \dot{N}_+(t-t_\mathrm{br})-n N_+(t)(\langle \sigma_\mathrm{a} v \rangle + \langle \sigma_\mathrm{r} v \rangle)
\end{equation}
with the condition that no positrons had been produced before the start time $t_s$: $\dot{N}_+(t<t_s)=0$.

\begin{figure}
    \centering
    \includegraphics[scale=0.8]{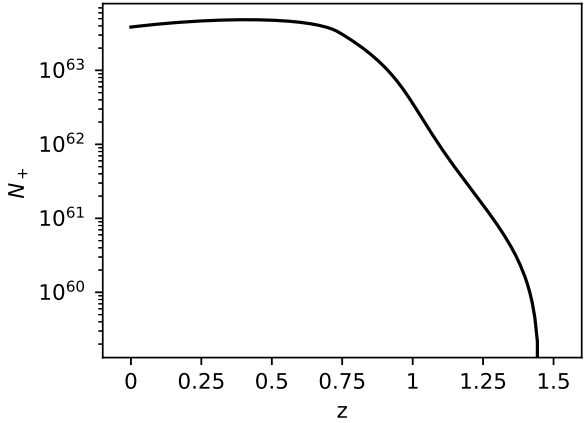}
    \caption{Evolution of the content of \textit{thermalized} positrons in the halo of  M31. Note that $z$ range is different from that in Figs.~\ref{fig:mdot} and \ref{fig:tracks}.}
    \label{fig:npos}
\end{figure}

The results are shown in Fig.~\ref{fig:npos}. Note that the thermalized positron content starts to grow at $z=1.44$. The time between $z=3$ and $z=1.44$ is 2.2~Gyr and equals the braking time for $T=10^6$~K, $\Gamma=10$.

Finally, to calculate the resulting 511~keV photon flux, we notice that each direct annihilation produces two photons and
so do one quarter of positronium annihilation (the remaining 3/4 decay into three continuum photons). The flux at the
Earth is
\begin{equation}
    F = \frac{2 n N_+(t_e) (\langle \sigma_\mathrm{a} v \rangle + \frac14\langle \sigma_\mathrm{r} v \rangle)}{4\pi d^2}
\end{equation}
where $d=750$~kpc is the assumed distance to M31. For the halo parameters $n=10^{-4}$~cm$^{-3}$, $T=10^6$~K, we obtain
$F=2.5 \times 10^{-4}$~photon~cm$^{-2}$s$^{-1}$. This is larger than what was obtained in \cite{Furlanetto2002} for the Virgo cluster and Cen~A ($\sim 10^{-6}$ and $\sim 10^{-5}$~photon~cm$^{-2}$s$^{-1}$ respectively.)

In order to estimate possible effects of faster braking due to advection we also considered the limiting case with $t_{br}=0$, i.e. instantaneous braking. It only slightly changed our results for $T=10^6$~K but for $T=10^5$~K the resulting flux estimate decreased almost two-fold. This behavior could be expected: in the former case the annihilation speed is low and we do not depend too much on the details of the immediate history of the SMBH evolution. In the latter case it is not true, now, with the decreased braking time we 'probe' more recent epoch of accretion when the accretion rate was considerably lower.

%%%%%%%%%%%%%%%%%%%%  DISCUSSION  %%%%%%%%%%%%%%%%%%%%%%%%
\section{Discussion}\label{sec:discussion}
The calculation of the SMBH growth history depends on several parameters and assumptions. First of all, we postulated
the values of the accretion efficiency $\varepsilon=0.1$ and Eddington ratio $\lambda=1$. The accretion efficiency is
theoretically bound between 0.054 for a non-rotating black hole and 0.42 for a maximally rotating black hole. In a number of studies of SMBH growth, including M04, this parameter was found in the range 0.05--0.4 (see \cite{Tucci2017} and references therein). Discussion of the $\lambda$ parameter in \cite{Tucci2017} shows that it can be roughly estimated as 0.01--0.1. In \cite{Marconi2004}, Marconi et al. compared the current mass function of the black holes which have hosted AGNs (they call them relic black holes) with the mass function of the SMBHs observed in the local Universe (they call them local black holes). In the calculations they adopted the values $\varepsilon=0.1$ and $\lambda=1$ because they provided the relic SMBH mass function
fairly close to that of the local SMBHs. One can think that the value of $\lambda$ is overestimated. However, one should bear in mind that the mass growth is assumed to take place when the AGN is active, i.e. radiates at the luminosity given by Eq.~\ref{eq:L}. The luminosity function is related to the SMBH mass function via Eq.~\ref{eq:phi} where there is a dependence on AGN duty cycle coming from $\delta(M,t)$ function. As Marconi et al. argue, the fact that the relic and local SMBH mass functions agree at $\lambda=1$ simply means that the growth takes place effectively when the AGN luminosity is close to the Eddington limit.

When calculating the SMBH growth history, we used the \textit{average} growth rate $\langle \Dot{M} \rangle$. Of course, for an individual object we do not have the precise growth history. However, our analysis shows that, if the annihilation time is long (e.g. as in the case $T=10^6$~K) then the positrons survive for a long time as well, hence the most important parameter is the final BH mass, because it determines the total mass accreted, the total energy radiated and the total positron population created. Even if the annihilation time is relatively short, our conclusions still hold if the main mass growth was at the same epoch, i.e. $z\approx1.2$ or later.

Another assumption vital for our reasoning is the existence of a hot gaseous halo around M31. Such hot halos are predicted by theories of galaxy formation \cite{White1978}, and currently, in a number of massive spirals they have been found \cite{Li2017}, \cite{Mirakhor2021}.
%\cite{Tuellmann2006} \cite{Li2007}
Discovery in X rays of a hot halo around M31 was also reported \cite{Otte2007}, and there is evidence from radio and UV data as well. In the work \cite{Westmeier2005} the authors observed high velocity clouds near M31, and two of these clouds demonstrated a 'head-tail' structure which can be attributed to the interaction with the ambient medium. Observations with Hubble Space Telescope in UV reveal the massive halo around M31 which includes hot components with $T\sim 10^5 - 10^6$~K \cite{Lehner2015}.

To investigate how the result depends on the parameters which are loosely constrained, we perform the calculations for several combinations of them. We try the halo temperature $T=10^5, 10^6$~K and the jet bulk Lorentz factor $\Gamma=5, 10, 20$. The resulting 511~keV fluxes are shown in Table~\ref{tab:results}.
\begin{table}[]
    \centering
    \begin{tabular}{c|ccc}
        \backslashbox{$T$,~K}{$\Gamma$} & 5 & 10 & 20 \\
         \hline
        $10^6$ & $4.6 \times 10^{-4}$ & $2.5 \times 10^{-4}$& $1.5 \times 10^{-4}$\\
        $10^5$ & $1.1 \times 10^{-4}$ & $8.7 \times 10^{-5}$ & $1.1 \times 10^{-4}$ \\
        \hline
    \end{tabular}
    \caption{Present time 511~keV photon fluxes at Earth in cm$^{-2}$s$^{-1}$ for various values of the halo temperature and the jet bulk Lorentz factor.}
    \label{tab:results}
\end{table}
For $T=10^6$~K both annihilation times exceed the Hubble time, therefore the flux is mostly affected by the total number of produced pairs. With the jet power fixed, the larger $\Gamma$ implies the smaller proton and pair production rates, hence the smaller resulting flux. However, for $T=10^5$~K positronium annihilation time is only 2~Gyr, while the braking time $t_\mathrm{br}$ for $\Gamma=10$ and 20 are respectively 2.2 and 4.4~Gyr and there an interesting interplay emerges: larger $\Gamma$  leads to longer braking times, meaning that more positrons survive since the epoch of the AGN luminosity peak. On the other hand, larger $\Gamma$ leads to lower rates of pair production (see eq. \ref{eq:ndot}). Consequently, the resulting photon flux depends on the bulk Lorentz factor non-monotonically which is evident from the second line of the table. 

\begin{figure}
    \centering
    \includegraphics[scale=0.8]{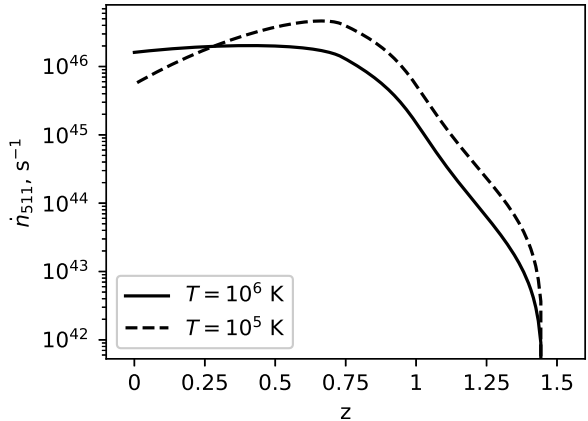}
    \caption{511~keV photon production rate in M31 halo for the halo temperature $T=10^6$ and $10^5$~K ($\Gamma=10$). Note that $z$ range is different from that in Figs.~\ref{fig:mdot} and \ref{fig:tracks}.}
    \label{fig:nphot}
\end{figure}

Finally, in Fig.~\ref{fig:nphot} we show the 511~keV photons production rate in the case $\Gamma=10$ for the two values of the halo temperature, $10^6$ and $10^5$~K. One can see that, as long as new pairs are created, the $T=10^5$~K medium produces more photons due to shorter annihilation time. When the accretion starts do decline and pair production drops, fast annihilation exhausts the positron population quicker. As a result, the present time photon production rate is larger when $T=10^6$~K.

The estimated flux can be compared to the observations. Siegert et al. \cite{Siegert2016} investigated the positron annihilation line in the Milky Way with INTEGRAL/SPI. Along with its diffuse emission they tried to model several point sources, such as Sgr~A*, Crab, Cyg~X-1. They also modeled M31 and found an upper limit of $10^{-4}$ photon cm$^{-2}$s$^{-1}$ at the significance level of $2\sigma$. The M31 galaxy was modeled as a \textit{point} source. In our setup, the emission originates from the halo, and even with the source size of 30~kpc, its angular size is $2^\circ$. So with the flux of  $\sim 10^{-4}$ cm$^{-2}$s$^{-1}$ it is tentatively close to the current limits and can be potentially detected with INTEGRAL or future MeV missions like e-ASTROGAM \cite{e-ASTROGAM}.

%%%%%%%%%%%%%%  CONCLUSIONS  %%%%%%%%%%%%%%%
\section{Conclusions}
AGN jets are a viable source of positrons in the Universe. If they are produced in AGN jets and are trapped in galactic gaseous halos, they can survive for substantial amount of time due to considerable braking time and long (for certain medium parameters) annihilation times. We calculated the positron production rate, their braking and annihilation times, and finally estimated the 511~keV photon flux at the present time due to the presumable past activity of the M31 nucleus, the source with the highest expected signal. To do so, we calculated the growth history of the SMBH in M31 following \cite{Marconi2004} and linked it to the AGN luminosity and positron production rate. We found that for a reasonable parameter combination, the present 511~keV photon flux at Earth can be as high as few times $10^{-4}$~cm$^{-2}$s$^{-1}$ and can be potentially observed in the near future.

\section*{Acknowledgements}
The authors thank Prof. Konstantin Postnov for fruitful discussions and the anonymous referees whose comments helped to improve the paper.
The work of the authors was supported by the Ministry of Science and Higher Education of Russian Federation
under the contract 075-15-2020-778 in the framework of the Large Scientific Projects program within the national
project "Science".  This research has made use of NASA's Astrophysics Data System.

\bibliographystyle{unsrt}

%\begin{center}
%   \bf{\large{Title.}}\\
%    M.S. Pshirkov${1,2,3}$, B.A. Nizamov$^1$ \\
%     \it{$^1$ Sternberg Astronomical Institute, Lomonosov Moscow State University, Moscow, 119234, Russia}\\
%    \it{$^2$ Institute for Nuclear Research of the Russian Academy of 	Sciences, Moscow, 117312, Russia}\\
%    \it{$^3$ Lebedev Physical Institute, Pushchino Radio Astronomy Observatory, 142290, Russia}\\
%
%\end{center}

%\section*{Abstract}
%In English

\end{document}